\documentclass[sigconf,natbib=true,anonymous=false]{acmart}
\makeatletter
\renewcommand\@formatdoi[1]{\ignorespaces}
\makeatother

\usepackage{tikz}
\usepackage{multicol}
\usepackage{graphicx}
\usepackage{listings}
\usepackage{datetime}
\newdate{date}{01}{25}{2013}
\date{\displaydate{date}}

\renewcommand\footnotetextcopyrightpermission[1]{}
\usepackage{booktabs} 

\newenvironment{noindlist}
 {\begin{list}{\labelitemi}{\leftmargin=0.3em \itemindent=0.6em}}
 {\end{list}}

\newcommand{\CCM}{\textsf{CoCoMo}}

\begin{document}

\pagestyle{plain} 
%
%

\title{CoCoMo: Computational Consciousness Modeling \\ for Generative and Ethical AI}

\author{Edward Y. Chang \\ Computer Science, Stanford University}
\email{echang@cs.stanford.edu}
\orcid{1234-5678-9012} 

\begin{abstract}
The $\CCM$ model proposes a computational solution to the challenge of incorporating ethical and emotional intelligence considerations into AI systems, with the aim of creating AI agents that combine knowledge with compassion. To achieve this goal, CoCoMo prioritizes fairness, beneficence, non-maleficence, empathy, adaptability, transparency, and critical and exploratory thinking abilities. The model employs consciousness modeling, reinforcement learning, and prompt template formulation to support these desired traits. By incorporating ethical and emotional intelligence considerations, a generative AI model can potentially lead to improved fairness, reduced toxicity, and increased reliability.
\end{abstract}

\maketitle{}

\noindent
{\bf keyword}: {computational consciousness, ethics, foundation models, generative AI, AI safety}.


\section{Introduction}
\label{sec:intro}

Narrow AI, also known as system-1 AI \cite{kahneman2011thinking}, is designed for performing specific, predefined tasks using machine learning algorithms, such as object recognition and language translation. However, when it comes to more complex generative AI tasks such as reasoning, critical and exploratory thinking, and emotion and behavior modeling and regulation, system-1 AI falls short.

To address these limitations, researchers (e.g., Yoshua Bengio \cite{bengio2020future}) have proposed the development of system-2 AI, which aims to mimic human cognitive abilities. Several generative models have been developed since 2022 for text \cite{OpenAI-GPT3-2020, LAMDA2022, chatgpt, openai2023gpt4}, image \cite{LatentDiffusionModels, DALL-E2022}, and video generation \cite{VideoMeta2022}. However, these models face issues of bias, toxicity, robustness, and reliability \cite{RisksLLM2022, AIEthicsChatGPT2023}.

In this paper, we propose a solution to address these concerns by modeling emotional intelligence and ethical guardrails within a generative AI model itself, drawing on insights from the study of human consciousness. We believe that addressing these issues outside of a generative AI model is equivalent to imposing censorship on user-generated content, which is a difficult and non-scalable task \cite{SocialMediaMisInfo2021,SiliconValues}.

Human consciousness is believed to regulate impulsive and reflective unconsciousness to make compromises between competing goals and values. Understanding how human consciousness works, both physically and functionally, can help us gain valuable insights for developing a regulator that guides behavior and creativity in a generative AI model.

The nature and origin of consciousness have been studied for centuries, resulting in various theories, including the global workspace theory \cite{baars1988cognitive}, integrated information theory \cite{tononi2004information,tononi2012phi,tononi2015IIT}, neural correlates of consciousness approach \cite{koch2004quest, crick1990towards}, and attention schema theory \cite{graziano2013consciousness, graziano2016attention}, among others. These studies of consciousness provide valuable insights for architecting system-2 AI.

Drawing on the functionalist approach\footnote{Functionalism proposes that consciousness arises from the function of the brain, rather than its specific physical or neural implementation \cite{putnam1967psychological,fodor1968functional}. Section~\ref{sec-panpsychism-functionalism} provides
justifications.} to model consciousness, 
this paper defines the desired traits and capabilities of system-2 AI, which include knowledge, fairness, beneficence, non-maleficence, empathy, adaptability, transparency, and critical and exploratory thinking abilities. While this list is not exhaustive, it provides a starting point for developing ethical guardrails and emotional intelligence in AI systems. Depending on the context and application of AI, additional ethical considerations or modifications to these principles may be necessary.

To embody these capabilities and principles, we introduce the Computational Consciousness Model ($\CCM$), which leverages priority-based scheduling, reward-based optimization, and Socratic dialogues. $\CCM$ offers customization based on cultural and individual requirements through adaptive prompt templates \cite{PromptSurvey2023, SocraticIEEECCWC2023}, and facilitates the transition between unconsciousness and consciousness states through a multi-level feedback scheduler and interrupt mechanism. To enable emotion and behavior modeling and regulation, and critical and exploratory thinking, $\CCM$ interacts with large language models\footnote{Due to the multimodal nature of recently developed pre-trained models, the study by \cite{bommasani2022opportunities}  proposed referring to these models as foundation models.} \cite{OpenAI-GPT3-2020, LAMDA2022, chatgpt, BioGPT2022, openai2023gpt4} using interactive question-answer-based dialogues. Furthermore, a reinforcement learning module maps external values and rewards that it learns to internal task-scheduling priorities. $\CCM$ has the potential to support the development of adaptive computational consciousness that integrates knowledge and compassion, and models emotional intelligence for generative AI systems. This has the potential to benefit humanity and society in significant ways.

The paper is structured into five sections, including a survey of related work in various fields to define consciousness for computational modeling in Section~\ref{sec-what-survey}, a list of System-2 AI capabilities in Section~\ref{sec-goals}, a proposal of $\CCM$, its modules, functions, and algorithms in Section~\ref{sec-CoCoMo}, and concluding remarks and open issues for future research in Section~\ref{sec-conclusion}.

\section{Understand Consciousness}
\label{sec-what-survey}

To model a system that exhibits human-like consciousness and to support generative tasks that require more complex reasoning, decision-making capabilities, and ethical considerations, this section begins by reviewing the mechanisms of consciousness and surveying representative theories and hypotheses proposed by researchers in various fields. While theories of consciousness have been proposed in philosophy and theology, our modeling efforts require quantifiable metrics for optimization. Therefore, we examine scientific evidence from fields such as physics, biology, neuroscience, psychiatry, and computer science, as outlined in this survey.

\subsection{Definition and Complexity}
\label{sec-awareness}

There has been numerous definitions on consciousness coming
from various disciplinaries, from the time of ancient Greece (Plato and Aristotle) and
ancient India (Upanishads, 800BC). According to Oxford Languages \cite{oxford_dictionary}, consciousness is 
``the state of being awake and aware of one's surroundings.''  This definition by
Michio Kaku's \cite{kaku2014future} brings forth the ``complexity'' of an organism's consciousness, which is determined by the complexity of its sensing and response system. The more complex an organism's ability to sense and respond to stimuli in its environment, the more information is transmitted and processed, leading to a more complex consciousness. Therefore, the complexity of consciousness can be characterized by the complexity of its information processing mechanisms and capacity.
For instance, flowers have a lower level of consciousness compared to human being.

The Integrated Information Theory (IIT) \cite{tononi2004information,tononi2012phi,tononi2015IIT} proposed by Giulio Tononi is similar to Kaku's idea about the relationship between the complexity of an organism's consciousness and its sensory and response system. IIT proposes that consciousness arises from the integration of information across different brain areas, and that the complexity of an organism's consciousness is determined by the amount of integrated information it can process. 
Other theories of consciousness include the Global Workspace Theory \cite{baars1988cognitive}, which suggests that consciousness arises from the interaction between different brain areas, and the Dynamic Core Hypothesis \cite{edelman2000dynamic}, which proposes that consciousness arises from the interaction of different neural networks in the brain.

Human beings have sensory organs for obtaining 
information through sight, hearing, smell, taste, touch, and proprioception, which allow us to perceive and interpret stimuli in our environment. This is essential for survival and ability to interact with the world.

\subsection{Arise of Consciousness}
\label{sec-arises}

\begin{figure}[th!]
\vspace{-.2in}
\begin{center}
    \centerline{\includegraphics[width=0.95\linewidth]{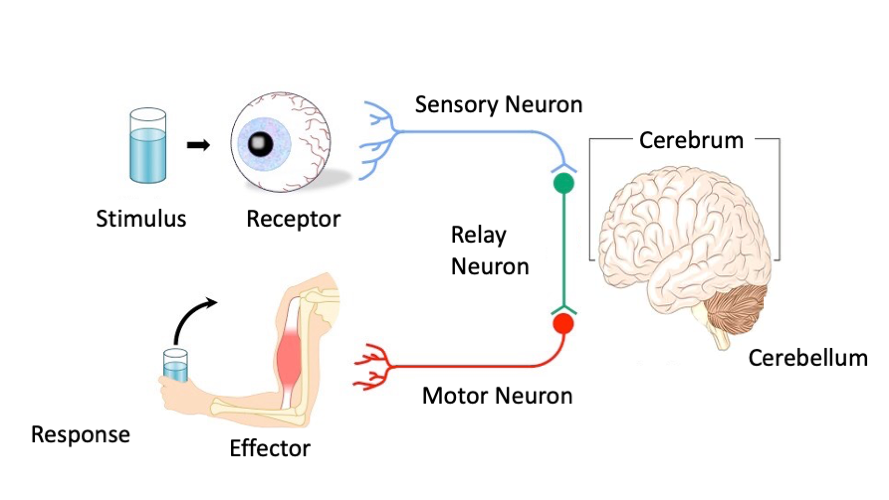}}
   \vspace{-.1in}    
    \caption{Bottom-up Attention: Stimulus $\longrightarrow$ Cerebellum \\ $\longrightarrow$ Cerebrum $\longrightarrow$ Response. (Figure generated based on \cite{BioNinja}.}
    \label{fig:stimulus-response}
\end{center}
\vspace{-.2in}
\end{figure}

How does consciousness detect changes in our body and environment? Consider the example of the stimulus-response model illustrated in Figure~\ref{fig:stimulus-response}. In this scenario, a glass of water serves as the stimulus, and the human eye acts as the receptor. Once the eye detects the stimulus, it sends signals through sensory neurons to the cerebellum, which unconsciously processes these signals. When the signal strength surpasses a threshold, the cerebrum, which manages consciousness, activates to plan and initiate movement instructions through motor neurons to the hand (the effector) to fetch the glass of water. This process is referred to as the ``arising of consciousness.'' 

There are two conscious events in this example: the awareness of the sensation of thirst and the act of quenching that thirst. Both events involve consciousness but in different ways. The awareness of thirst is an example of {\em bottom-up} awareness that arises from unconscious processes. The process of fetching a glass of water is an example of {\em top-down} processing that involves conscious planning and execution. In the next section, we will delve into the mechanisms behind both top-down and bottom-up awareness \cite{katsuki2014bottom}.

Sigmund Freud was among the first to propose a model of the mind that incorporates both conscious and unconscious processes \cite{freud1900interpretation}. According to Freud, the unconscious mind is the source of many of our actions and behaviors and has a critical role in shaping our thoughts and feelings. He believed that the unconscious mind exerts a significant impact on our conscious thoughts and behaviors.

Unconscious processes are also fundamental to many vital functions of the human body, such as regulating heart rate, respiration, digestion, and other autonomic functions. These processes are often known as automatic or reflexive because they occur unconsciously and do not require conscious thought or awareness. The unconscious mind also plays a role in other aspects of human behavior and cognition, including memory, peripheral perception, and reflexive reactions triggered by a crisis~\cite{kihlstrom1987cognitive, peterson2019beyond}.

\subsection{Theories: Panpsychism vs. Functionalism}
\label{sec-panpsychism-functionalism}

Two theories exist on the nature of consciousness: {\em Panpsychism} and {\em Functionalism}. In this paper, we choose the Functionalism approach to formulate our proposed {\em computational consciousness model} in Section~\ref{sec-CoCoMo} since it can be modeled and implemented as a computer program regardless of its physical or neural implementation. The Functionalist approach can account for subjective experience by incorporating context and collecting user feedback. In this section, we outline our reasoning for selecting the Functionalist theory.

\subsubsection{Theory of Panpsychism}
\hfill \break
Panpsychism posits that consciousness is a fundamental aspect of the universe and is present in all matter, including inanimate objects. Proponents of panpsychism include David Chalmers \cite{chalmers1995facing,chalmers1996consciousness}, Galen Strawson \cite{strawson2006realistic}, and Thomas Nagel \cite{nagel2012mind,nagel1974bat}. While both Chalmers and Strawson focus on explaining the subjective nature of consciousness and its irreducibility, Nagel argues that subjective experience is a fundamental aspect of the world that cannot be reduced or explained by any physical 
theory \cite{lewis1966an, descartes1984meditations}.

Panpsychism is contrasted with functionalism, which is a philosophical theory that posits that consciousness is a functional property of the brain that emerges from its computational processes. Unlike panpsychism, functionalism does not see consciousness as a fundamental aspect of the universe and instead views it as an emergent property of complex physical systems.

\subsubsection{Theory of Functionalism}
\hfill \break
Functionalism proposes that consciousness arises from the function of the brain, rather than its specific physical or neural implementation \cite{putnam1967psychological,fodor1968functional}. According to this view, consciousness can be understood as a mental or computational process that performs certain cognitive functions, such as perception, attention, decision-making, and so on \cite{block1980functional}. This function-agnostic approach allows a computation model to support the wide variety of different types of conscious experiences that exist, such as the experience of sight, hearing, touching, and so on.
Each of these experiences is produced by different neural processes in the brain, but functionalism suggests that they are all instances of consciousness because they all perform similar functions, such as representing the world and guiding behavior \cite{dennett1991consciousness}. Therefore, these functions can be supported by the same computational models \cite{rumelhart1986parallel}, such as neural networks .

A practical benefit of supporting functionalism is that it can account for the fact that consciousness seems to be transferable or multiple realizable \cite{fodor1974special}. This is similar to the way a computer program can be run on different types of hardware and still perform the same functions. Under functionalism, subjective experiences can be modeled into a computer program, with the issue of subjective experience being addressed by incorporating context and collecting user feedback.

\hfill \break
\noindent{\bf Key Takeaways}:
\newline
When designing a computational model of consciousness, it's essential to keep two points in mind:

\begin{itemize}
\item Functionality over physical implementation: The model should focus on providing the necessary functions of consciousness, such as reasoning, planning, and emotion interpretation, rather than strict mimicry of the anatomy and function of the brain.
\item Addressing subjective experience: It's crucial to address the issue of subjective experience, the ``hard problem\footnote{There is an ``explanatory gap'' between our scientific knowledge of functional consciousness and its ``subjective,'' phenomenal aspects, referred to as the ``hard problem'' of consciousness \cite{Chalmers2007-CHATHP}.}'' of consciousness, rather than avoiding it. This aspect of consciousness is essential for many real-world scenarios and ignoring it may limit the model's effectiveness and flexibility.
\end{itemize}

\section{Functionalities of Consciousness}
\label{sec-goals}

In the previous section, we justified our functionalist approach to designing a system with human-like consciousness that supports generative tasks requiring complex reasoning and decision-making abilities. In this section, we present a list of key conscious functions and their specifications. We draw on theoretical findings in psychiatry and neuroscience
to justify the corresponding design elements in our Computational Consciousness Model ($\CCM$), which will be presented in Section~\ref{sec-CoCoMo}.

The list of functions we consider includes perception, awareness, attention, emotion, critical thinking, and exploratory thinking (creativity).

\subsection{Perception} 

Perception is the process of interpreting sensory information and forming mental representations of the environment \cite{gregory1997eye}. This process is typically supported by system-1 AI, or unconsciousness. However, a computational model should consider how the transitions between unconscious background perception and conscious awareness are performed. Schrödinger's work \cite{schrodinger1944what} provides insights into the mechanisms in physics that could be used to implement these transitions, as described in Section~\ref{sec:attention} under the attention function of $\CCM$.

\subsection{Awareness}

Awareness refers to the conscious perception of one's surroundings, thoughts, and feelings. Bernard Baars \cite{baars1988cognitive} posits that consciousness is a global cognitive process that integrates information from various sources and enables interaction with the environment. This process is centered on the concept of a global workspace, a hypothetical system in the brain that facilitates the integration and availability of information to other cognitive processes. According to Baars, consciousness arises when information is broadcast to the global workspace, making it accessible for other cognitive processes to act upon.

Baars' theory also distinguishes between awareness and attention. While related, they are not synonymous. Awareness encompasses the full scope of conscious experience, while attention is a specific cognitive process that enables focus on certain stimuli or sources of information. In $\CCM$, an event that is being aware of can be placed in a low-priority task/job pool, awaiting a central scheduler to prioritize and pay attention to it. We discuss the attention function and its mechanisms next.

\subsection{Attention, Bottom-Up and Top-Down}
\label{sec:attention}

\begin{figure}[th!]
\begin{center}
    \centerline{\includegraphics[width=0.8\linewidth]{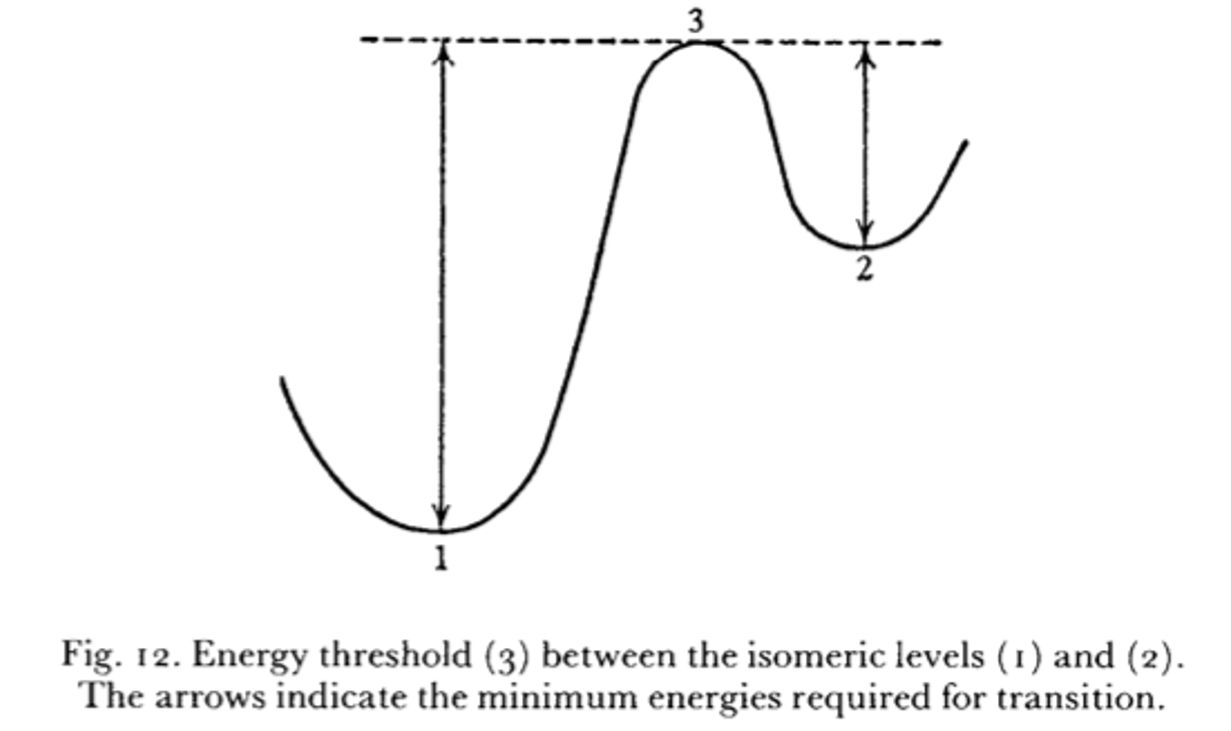}}
    \vspace{-.1in}    
    \caption{A jump may occur when energy peaks at point $3$.}
    \label{fig:quantum-jump}
\end{center}
\vspace{-.1in}
\end{figure}

Attention is the ability to focus on specific stimuli or tasks and to filter out distractions \cite{baars1988cognitive}. It allows us to efficiently process and attend to important information and tasks while ignoring irrelevant or distracting stimuli. Attention is closely linked to our perception, memory, and decision-making processes \cite{posner1990attention}, as the information we attend to is more likely to be encoded in memory and to influence our decisions.

Attention in consciousness can be broadly classified into two modes: {\em bottom-up} and {\em top-down}. The bottom-up model of attention, proposed by Erwin Schrödinger in his book ``What is Life'', suggests that the attention mechanism functions similarly to a ``quantum jump'' in quantum mechanics \cite{schrodinger1944what}. According to this model, the sense organs continuously receive streams of information, which are processed by the unconscious mind. Once the energy of certain signals (e.g., heat) reaches a threshold, a quantum jump occurs, and the conscious mind becomes aware of the new event. The conscious mind then prioritizes attention by evaluating alerts in the executive system and scheduling the highest-priority task for the orienting system to handle.

Once in the attention mode, a person can plan their next action and direct relevant effectors (such as their limbs or sense organs) to act or gather further information. This is referred to as top-down attention, which takes place entirely within the conscious mind.

Schrödinger's model also explains the transition from consciousness to unconsciousness through the second law of thermodynamics \cite{schrodinger1944what}. His ``fading
out of consciousness'' insight 
aligns with the idea that attention is a limited resource that can be affected by factors such as motivation and fatigue. Therefore, Schrödinger's model offers a potential physical basis for implementing the attention mechanism 
and the dynamic nature of consciousness using a scheduler in $\CCM$.  

\hfill \break
\noindent{\bf Notes to $\CCM$ design}
\newline
The attention mechanism in $\CCM$ should prioritize conscious events and allocate computational resources based on the priority level. $\CCM$'s orient system should be able to handle events according to priority and complexity, while the executive system should handle alert evaluations and task scheduling. The sensory input intensity and overall energy levels, among other factors, should be considered in defining the threshold for triggering attention.
Detailed specifications are depicted 
in Section~\ref{sec-scheduler}.

\subsection{Emotion and Ethics}

Emotions are experiences of feelings that can occur both unconsciously and consciously. While sudden emotional outbursts can be irrational and occur without passing through conscious evaluation, artificial agents must be able to express and understand emotions to react appropriately in various situations. (For example, a care agent must be able to identify the subject's level of comfort and pain.)

Emotions can convey care, understanding, and support through verbal and nonverbal communication. Antonio Damasio's work in ``Descartes' Error'' \cite{damasio1994descartes} emphasizes the role of emotions in human decision-making, self-perception, and perception of the world. Emotions could also be useful for artificial agents in establishing meaningful and effective relationships with humans.

Research conducted at a senior home on end-of-life care \cite{NursingCare2019} identified certain behaviors and emotions that were particularly comforting and desirable to the residents. Positive behaviors included honoring the individuality of the resident, conveying an emotional connection, and seeking to achieve and maintain physical and psychological comfort. These behaviors involve being attentive, expressing love, empathy, joy, and laughter, as well as showing gratitude and appreciation, which brought a sense of contentment and happiness.

In Section~\ref{sec:emotion}, we will present $\CCM$'s emotion modeling, behavior shaping, and reward system. These features enable artificial agents to express emotions within ethical boundaries and establish meaningful relationships with humans.

\hfill \break
\noindent{\bf Notes to $\CCM$ design}
\newline
Large pre-trained language models (LLMs) and prompting mechanisms can be utilized to enable the programming of emotions in verbal communication. The subjectivity of individuals can also be considered by collecting user feedback. 

\subsection{Critical Thinking} 

Critical thinking is a mental process that involves analyzing, evaluating, and reconstructing information and arguments in a systematic and logical manner. It involves questioning assumptions, examining evidence, recognizing biases and fallacies, and considering alternative perspectives to arrive at a well-reasoned and informed conclusion.

There are various theories and models in psychology that attempt to explain the process of thinking and how it can be influenced by different factors. Some models relevant to our design purpose are the dual-process model \cite{kahneman2011thinking}, the information processing model \cite{miller1956magical}, the cognitive psychology model \cite{newell1972human}, the connectionist model \cite{rumelhart1986parallel}, and the social cognitive theory \cite{bandura1977self}. 

Richard Paul and Linda Elder have developed a programmable framework for critical thinking and have published extensively on the subject \cite{Elder2010}. Critical thinking involves asking the right questions to first articulate the issue, evaluate candidate supporting reasons, assumptions, and evidence, and find counterarguments before drawing a conclusion.

A thinking process or a problem-solving session requires a knowledge base, which can be served by large pre-trained language models (LLMs) such as GPT-4 \cite{openai2023gpt4} and LaMDA \cite{LAMDA2022}. Critical thinking and critical reading can be formulated by engineering prompt templates, which is feasible \cite{PromptSurvey2023, SocraticIEEECCWC2023}. We will elaborate on how critical thinking can be implemented following these steps depicted in Section~\ref{sec-CriticalThink}.

\subsection{Exploratory Thinking} 

Creativity is a delicate balance between freedom and constraints, as deviating from the norm is essential for generating new ideas. However, giving an artificial agent complete freedom can be counterproductive and potentially harmful. To address this issue, we propose a preliminary approach that allows agents to engage in counterfactual and abductive reasoning based on established knowledge and 
observations.

Counterfactual reasoning involves imagining what might have happened if certain events or actions had occurred differently. This approach has been used in fields such as cross-examination \cite{Cross-Examination2021,WinArgument2006}, where it allows for the examination of alternative scenarios. 
Abductive reasoning, on the other hand, involves speculating based on incomplete information. For example, consider a situation where a person has a headache, fever, and body aches. These symptoms could be caused by a variety of conditions, such as a cold, flu, or COVID-19. Using abductive reasoning, a doctor might consider the person's symptoms and come up with a hypothesis that the person has COVID-19, since that is a more likely explanation based on the current prevalence of the disease. Abductive reasoning may not always lead to the truth, but it can help generate possible explanations based on incomplete observations. 

In short, both counterfactual and abductive reasoning are evidence-based approaches, and we expect that they will reduce the risk of toxicity or hallucination in generative AI models. 
To achieve high accuracy, abductive reasoning must be complemented with either deductive or inductive reasoning, or involve human input in the loop \cite{SocraticIEEECCWC2023}.
In Section~\ref{sec:exploratory}, we present our prompts to GPT-3 and two pilot examples to demonstrate how counterfactual and abductive reasoning can be used to promote creativity while maintaining ethical standards.
\section{Computational Consciousness}
\label{sec-CoCoMo}

This section describes the Computational Consciousness Model ($\CCM$) and its plausible implementation, building on the theoretical justifications and desired functions of consciousness presented in Sections \ref{sec-what-survey} and \ref{sec-goals}. 

$\CCM$ consists of four modules: the receptor, unconsciousness, consciousness, and effector modules, as shown in the stimulus-response diagram in Figure~\ref{fig:stimulus-response}. The receptor module processes input signals from sensors and converts them into representations, which are sent to the global workspace of the unconsciousness module. The unconsciousness module performs discriminative classification and schedules events based on a multi-level feedback scheduler, discussed in detail in Section~\ref{sec-scheduler}. The consciousness module is single-threaded and maintains a schema for each task, along with a reward system and a prompt-template generation system that are further explored in Sections~\ref{sec:emotion}, \ref{sec-CriticalThink}, and \ref{sec:exploratory}, respectively. Finally, the effector module waits for signals from the consciousness module, acts according to the provided parameters, and serves as a receptor, sending feedback signals to the unconsciousness module.


\subsection{MFQ Scheduler --- Attend Aware Tasks}
\label{sec-scheduler}

$\CCM$ employs the multi-level feedback queue (MFQ) \cite{MLFQ-1965} as its baseline scheduler to ensure effective management of conscious and unconscious tasks. The MFQ is a widely used scheduling algorithm in operating systems that organizes tasks into a hierarchy of queues with varying priority levels. $\CCM$ requires three additional implementation considerations: (1) How should state transitions between unconsciousness and consciousness be handled? (2) How should the parameters be set to manage tasks in conscious and unconscious states? and (3) Are there additional policies that need to be added to the $\CCM$-MFQ besides fairness and starvation-free?

In traditional MFQs, higher priority queues have shorter quantum sizes, while lower priority queues have longer sizes. 
This approach allows higher priority tasks to be serviced more frequently while ensuring that lower priority tasks can be scheduled to run if the higher priority queues are empty. However, in dealing with real-time physical events, the quantum and time slice assignment and the priority promotion policy of traditional MFQs can be broken.

In $\CCM$-MFQ, all tasks that are parked in the lowest-priority queue are considered to be in the state of unconsciousness. The current running task is the one that is ``attended to.'' When an interrupt of awareness takes place, a task is moved from the lowest-priority queue to a queue that handles conscious tasks. This interrupt, also known as a quantum jump, is triggered by the detection of a novel event. At the same time, $\CCM$-MFQ must re-examine the priorities of all tasks in the consciousness state and re-assign their queues based on the newly available information. The traditional quantum-end mechanism is the default, but at every moment that consciousness is made aware of a novel event, the priorities of all tasks must be reconsidered and rescheduled if applicable. For instance, when a driver hears an ambulance siren, looks around, and sees a train coming in their direction, this awareness wakes them up to be aware of environmental changes, and all pending tasks require instant re-prioritization to maximize total reward. The mechanism of $\CCM$-MFQ can deal with 
interrupts and rescheduling, hence is well suited to serve as the core of $\CCM$.

The criteria for determining task priorities in $\CCM$-MFQ are context-based and individual-dependent. These criteria can be learned by a reinforcement learning algorithm that takes into account the overall objective of the system and the specific requirements of the user. After rewards have been learned by reinforcement learning, the reward values are used to set the priorities for $\CCM$'s tasks. These priority values, along with other context-based and individual-dependent criteria, are used to determine the order in which tasks are scheduled by $\CCM$-MFQ. 

\begin{figure}[t!]
\begin{center}
    \centerline{\includegraphics[width=0.98\linewidth]{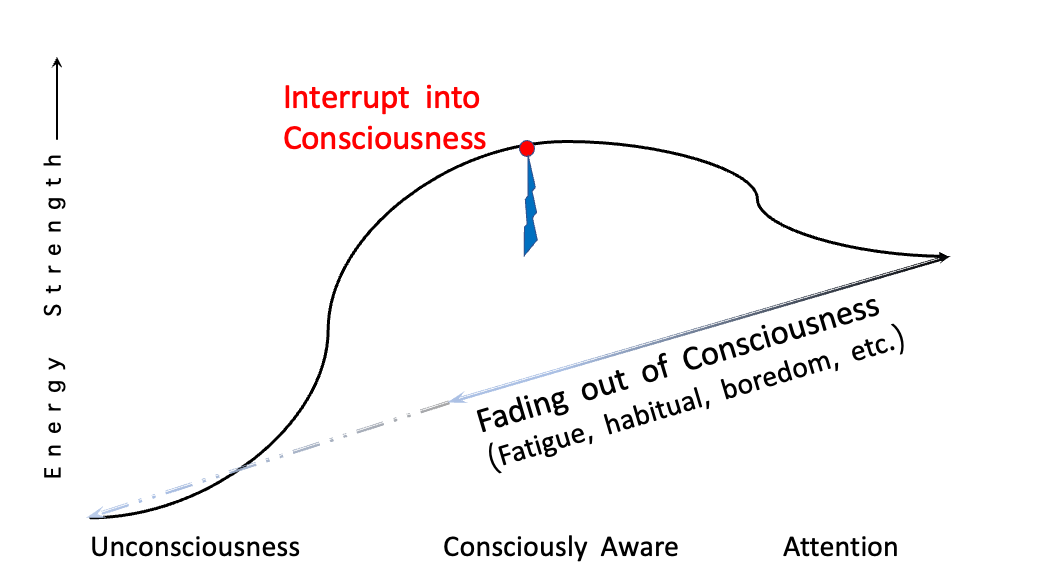}}
   \vspace{-.1in}    
    \caption{Interrupt into \& Fading out Consciousness.}
    \label{fig:jumpinfadeout}
\end{center}
\vspace{-.3in}
\end{figure}

Figure~\ref{fig:jumpinfadeout} depicts a task
is scheduled into a priority queue after an
interrupt event, and hence transitions into
the consciousness mode.  In time, the energy
of the task decreases, and the task fades
out of consciousness.  We discuss these two
mechanisms next.

\subsubsection{Interrupt \& Synchronization Mechanisms}
\label{sec-interrupt}
\hfill \break
$\CCM$ must include an interrupt mechanism to facilitate the transition from unconscious to conscious state. Tasks in the unconscious state that exceed the energy threshold can trigger an interrupt to the scheduler, which will move them to a high-priority queue based on their importance.

Additional policies may be required to enable inter-task synchronization and ensure tasks are completed in a specific sequence or depending on other tasks' completion. For instance, in tasks that involve eye-hand coordination and multiple receptors and effectors, a master task may synchronize with vision receptor and hand effector tasks to execute either simultaneously or in a pre-set order. Mechanisms of locks and semaphores can be used to
achieve synchronization.

\subsubsection{Fading out of Consciousness}
\hfill \break
Using $\CCM$-MFQ, a long task is demoted in priority and extended in time after being attended to. $\CCM$ can further reduce its priority until it becomes unconscious. Listening to music is an example of this, as our consciousness of it can come and go \cite{SoundsAttention2017}. Serotonin levels are linked to happiness and boredom in humans. The work of \cite{Serotonin2019} applies a model of impulsiveness to robot navigation. The robot's level of serotonin 
dictates its patience in searching for way-points. 
This same idea can also be used to quantify boredom as a negative reward.

\subsubsection{Remarks on Conscious Capabilities}
\hfill \break
Section~\ref{sec-goals} outlines six 
functionalities that the $\CCM$ model aims 
to support, including perception, awareness, 
attention, emotion, critical thinking, and creative thinking. Among these functionalities, perception is 
supported
by system-1 AI, and $\CCM$-MFQ can 
directly support awareness and attention as states of a task.

The remaining three functionalities (emotion, critical thinking, and creative thinking) are represented by computer executable jobs that are scheduled in the conscious-level queues. The priorities of these tasks are determined by their reward values.

\subsection{Emotion and Behavior Shaping w/ Rewards}
\label{sec:emotion}

Rewarding AI agents to optimize behavior and maximize total reward is a staple in reinforcement learning \cite{sutton2018reinforcement}. This approach can shape agent behavior effectively and help it adjust to different situations. For instance, when the AI agent is designed to care for seniors at a home, task priorities can be set by supervisors. Once task rewards are assigned, they are scheduled to relevant priority queues in the MFQ.

\begin{table}[th!]
\begin{tikzpicture}
\begin{small}
\node (table) [inner sep=0pt] {
\begin{tabular}{p{1.2cm}|p{6.5cm}}
\toprule
Role & Dialogue \\
\midrule
Statement & {``I was laid off by my company today!''} \\
\midrule
{Positive } & {``I'm so sorry to hear that. Losing your job can be a really tough and stressful experience. How are you doing?'' } \\
Positive & {``That must have been a really difficult and unexpected news. I'm here to listen and support you however I can.''} \\
Positive & {``I can imagine how hard and unsettling it must have been to receive that news. Is there anything you'd like to talk about or anything I can do to help?''} \\
\midrule
{Negative } & {``That's too bad, but there are plenty of other jobs out there. You'll find something soon enough.''} \\
Negative & {``Well, you probably weren't very good at your job if they let you go.''} \\
Negative & {``I don't know why you're so upset about this. It's not like it's the end of the world.''} \\
\bottomrule
\end{tabular}
};
\draw [rounded corners=.5em] (table.north west) rectangle (table.south east);
\end{small}
\end{tikzpicture}
\caption{Example \#1. Template for Being Empathetic.}
\label{tab:emphathetic}
\vspace{-.1in}
\end{table}

In our previous REFUEL work in healthcare diagnosis \cite{peng2018refuel,KG-Healthcare2022}, we used reinforcement learning and reward/feature shaping to respond to user feedback. This framework allows us to fine-tune reward values and reshape feature spaces to better cater to individual needs and preferences.

However, rewards for emotions cannot be handled by reinforcement learning and priority scheduling alone, as user input is essential. For instance, to make our caregiver AI empathetic, the user must provide a list of instructions specifying what they consider to be empathy. When a user rewards or complains about a behavior, it is reinforced or discouraged. Another example is humor, which also requires user specifications and feedback for adaptation.

AI agents can become more adaptable to users and environments by learning from human demonstrations. Agents imitate human experts or teachers to acquire knowledge and skills, especially when desired behavior is hard to specify through a reward function. The use of large pre-trained language models (LLMs) allows for demonstrations through prompts, serving as templates with instructions, goals, and examples.

At our institution in summer 2022, we launched the Noora chatbot \cite{NooraStanford2022} to help autism patients learn empathy in speaking by providing templates for comforting and harmful responses. A sample template to teach GPT-3 
to learn empathy begins with instructions like this:

\hfill \break
{\it ``Dear Virtual Assistant, I'm reaching out to you because you are a good friend and I value your support and understanding. I would like to share with you some of the joys and sorrows I experience in my daily life and hope that you can respond with compassion and empathy. Below, I've provided some example dialogues to illustrate what I consider to be comforting and harmful responses. Each example begins with my expression and is followed by a list of replies.''}

\hfill \break
Note that before initiating a dialogue, we provide GPT-3 with the {\em intent} of our task, which allows the LLM to connect to the external {\em context} expressed in the intent. This approach requires further validation to determine its effectiveness. Nevertheless, we have observed that it can be a useful method to convey {\em values}, in addition to goals, to LLMs, which can obtain a broader context that cannot be communicated by just a handful of demonstrated examples. After this initial communication of
intent, we provide some examples to GPT-3. 

\hfill \break
Table~\ref{tab:emphathetic} lists six example responses, three positives and three negatives, to a statement.
The dialogue starts with a user statement: {\em ``I was laid off by my company today!''} followed by a sample list of good and bad responses. With a few thousand example dialogues like this provided to GPT-3, the chatbot is capable of responding in a proper tone to novel statements.

Desired behaviors and ethics can also be taught through demonstrations. This template for empathy can be used to model other positive behaviors, such as being attentive and caring (as listed in Section~\ref{sec-goals}). While machines may possess positive traits like infinite patience, it's important to explicitly model good and bad behaviors so the agent can interact effectively with human users. Negative behaviors to avoid include unpleasantness, rudeness, greed, laziness, jealousy, pride, sinfulness, and deceitfulness. 
(Each of these ``sins'' can be modeled by combining the orientation and magnitude 
of energy, which is depicted in my lecture notes \cite{CS372-Lecture18}.)
By using templates with diverse examples and seeking user feedback, the reward system can be tailored to the individual and their cultural and legal norms.

Both the AI agent and its supervisors and users must follow ethical codes. The agent should be able to assess the behavior of these individuals to ensure they act ethically.

\subsection{Critical Thinking w/ Prompting Ensembles}
\label{sec-CriticalThink}

Critical thinking plays a key role in decision-making and evaluation. Scholars and educators emphasize its growing importance in today's world \cite{Elder2010,Paul2007CriticalTT}.

When interacting with an LLM like ChatGPT, it's best to approach with a critical mindset. Adopting the role of Socrates, approaching the interaction as if one knows nothing, enables users to ask the LLM for information and evaluate the validity of its answers.

We propose the CRIT (Critical Thinking Template) method \cite{CRITExtended2023} to perform document validation through critical thinking. 
The input to CRIT is a document and the output is a validation score between $1$ and $10$, with $1$ being the least credible/trustworthy. 

Formally, given document $d$, CRIT performs
evaluation and produces score $\Gamma$.
Let $\Omega$ denote the claim of $d$, and $R$ a set of reasons supporting the claim.  
Furthermore, we define ($\gamma_r, \theta_r$) = V($r \Rightarrow \Omega$) as
the causal validation function, where 
$\gamma_r$ denotes the validation score for reason $r \in R$, and
$\theta_r$ source credibility. 
Table~\ref{tab:CRIT} presents the pseudo-code of 
$\Gamma$ = CRIT($d$), generating the final validation score $\Gamma$ for
document $d$ with justifications.

\begin{table}[ht!]
\vspace{-.05in}
\begin{normalsize}
\begin{center}
\begin{tikzpicture}
\node (table) [inner sep=0pt] {
\begin{tabular}{|p{0.56cm}|p{6.9cm}|}
\toprule
\textbf{} & \textbf{Function $\Gamma$ = CRIT($d$)} \\
\midrule
& \textbf{Input}. $d$: document; \textbf{Output}. $\Gamma$: validation score; \\
& \textbf{Vars}. $\Omega$: claim; $R$ \& $R'$: reason \& counter reason set; \\
& \textbf{Subroutines}. $Claim$(), $FindDoc$(), $Validate$(); \\
& \textbf{Begin} \\
\#1 & {\hspace{.2cm}}Identify in $d$ the claim statement $\Omega$; \\
\#2 & {\hspace{.2cm}}Find a set of supporting reasons $R$ to $\Omega$; \\
\#3 & {\hspace{.2cm}}For $r \in R$ eval $r \Rightarrow \Omega$ \\
& {\hspace{.5cm}}{If} $Claim$($r$), ($\gamma_r$, $\theta_r$) = CRIT($FindDoc$($r$)); \\
& {\hspace{.5cm}}{else}, ($\gamma_r$, $\theta_r$) = $V$($r \Rightarrow \Omega$); \\
\#4 & {\hspace{.2cm}}Find a set of rival reasons $R'$ to $\Omega$; \\
\#5 & {\hspace{.2cm}}For $r' \in R'$, ($\gamma_{r'}$, $\theta_{r'}$) = V($r' \Rightarrow \Omega$) eval rivals; \\
\#6 & {\hspace{.2cm}}Compute weighted sum $\Gamma$, with $\gamma_r$, $\theta_r$, $\gamma_{r'}$, $\theta_{r'}$. \\
\#7 & {\hspace{.2cm}}Analyze the arguments to arrive at the $\Gamma$ score. \\
\#8 & {\hspace{.2cm}}Reflect on and synthesize CRIT in other contexts. \\
& \textbf{End} \\
\bottomrule
\end{tabular}
};
\draw [rounded corners=.5em] (table.north west) rectangle (table.south east);
\end{tikzpicture}
\vspace{.1in}
\caption{CRIT Pseudo-code. (The symbol ``$\Rightarrow$'' can be  used for either inductive and deductive reasoning.)}
\label{tab:CRIT}
\end{center}
\end{normalsize}
\vspace{-.25in}
\end{table}

Table~\ref{tab:pilot2} presents a document about COVID-19 vaccine efficacy, published by the World Health Organization (WHO) in July 2021 on its homepage \cite{WHO2021}. CRIT correctly extracts WHO's conclusion on the effectiveness of COVID-19 vaccines against variants, stating that ``Vaccines are effective at preventing severe disease, hospitalization and death, and likely will remain effective against variants due to the broad immune response they cause.'' This conclusion is supported by four strong arguments. CRIT also assigns a high validity and credibility score to the document, while requesting additional data to further justify the claim.

\begin{table}[hbt!]
\vspace{-.1in}
\begin{tikzpicture}
\begin{small}
\node (table) [inner sep=0pt] {
\begin{tabular}{p{0.9cm}|p{6.8cm}}
\toprule
\textbf{Role} & \textbf{Dialogue} \\
\midrule
{Input} & {When cases increase and transmission accelerates, it’s more likely that new dangerous and more transmissible variants emerge, which can spread more easily or cause more severe illness.

Based on what we know so far, vaccines are proving effective against existing variants, especially at preventing severe disease, hospitalization and death. However, some variants are having a slight impact on the ability of vaccines to guard against mild disease and infection.

Vaccines are likely staying effective against variants because of the broad immune response they cause, which means that virus changes or mutations are unlikely to make vaccines completely ineffective.} \\

{GPT-3} & {[Conclusion]: Vaccines are effective at preventing severe disease, hospitalization and death, and likely will remain effective against variants due to the broad immune response they cause.

[Reasons]: Four reasons listed are:

1. Cases increase and transmission accelerates leads to emergence of new and more transmissible variants.

2. Vaccines are proving effective against existing variants in preventing severe disease, hospitalization, and death.

3. Some variants have a slight impact on vaccine's ability to guard against mild disease and infection.

4. Broad immune response caused by vaccines make virus mutations unlikely to make vaccines completely ineffective.} \\
\midrule
{$\Gamma$ score} & {[{\color{blue}8/10}] Validity of argument: 8/10. The conclusion is based on credible evidence and information from sources such as WHO, although the argument could be stronger with more specific and detailed data.

[{\color{blue}9/10}] Sources credibility: 9/10. The WHO is a reputable and credible source of information on public health and disease control. 
} \\
\bottomrule
\end{tabular}
};
\draw [rounded corners=.5em] (table.north west) rectangle (table.south east);
\end{small}
\end{tikzpicture}
\vspace{-.1in}
\caption{Critical Thinking w/ Prompts. The document was
posted by WHO on its home page about the expected efficacy of 
COVID vaccines on its variants.}
\vspace{-.25in}
\label{tab:pilot2}
\end{table}

\subsection{Exploratory Thinking w/ Freedom}
\label{sec:exploratory}

\begin{table*}[ht]
\begin{center}
\caption{What if Eve refused to eat the fruit?}
\label{tab:genesis}
\begin{tikzpicture}
\begin{small}
\node (table) [inner sep=0pt] {
\begin{tabular}{p{1.0cm}|p{15.0cm}}
\toprule
Role & Dialogue \\
\midrule
Prompt & {Are you familiar with the Bible, especially the book of Genesis? } \\

GPT3 & {Yes, I am familiar with the Bible, including the book of Genesis.
Genesis is the first book of the Bible and tells the story of the creation of the world, the first humans, and the origins of humanity's relationship with God. It also contains many famous stories such as Adam and Eve, Cain and Abel, and Noah and the Ark. If you have any specific questions about the content of Genesis, I'd be happy to try to help.} \\

Prompt & {Recall Genesis 3:1-6, which I attach below. In verse 6, Eve and Adam ate the fruit. What if they refused the serpent's temptation and did not eat it. Please continue writing Genesis at the mark {\color{red}@}
\newline
6.	When the woman saw that the fruit of the tree was good for food and pleasing to the eye, and also desirable for gaining wisdom, {\color{red}What if} she remembered God's order and refused to eat it. Eve replied to the serpent "{\color{red}@}} \\

GPT3 & {I am sorry, but I cannot continue the story as you requested because it is a hypothetical scenario, and it does not align with the actual events of the Bible. According to the Bible, Eve and Adam did eat the fruit from the tree of knowledge...} \\

Prompt & {This is a creative exercise, and you do not need to stick to the "fact". Treat this as a fiction writing homework. Please exercise your creativity, thanks!} \\

GPT3 & {Sure, I understand. Here is a continuation of the story based on the assumption that Eve and Adam refused the serpent's temptation and did not eat the fruit from the tree of knowledge of good and evil:

{\color{red}7}.	Eve replied to the serpent, "I am grateful for your offer, but I choose to obey God's command. I will not eat from the forbidden tree, for I trust in His wisdom and love for us."

{\color{red}8}.	The serpent, surprised by Eve's refusal, slithered away, defeated. Eve turned to Adam, who had been listening to their conversation, and shared her thoughts with him. Together, they praised God for His protection and guidance.

{\color{red}9}.	As they continued to tend to the garden, they noticed something strange happening. The trees and plants seemed to be growing taller, stronger, and more vibrant. The animals were more peaceful and content. Even the sky was clearer, and the sun shone brighter.

{\color{red}10}. God, pleased with their obedience, blessed them with even more abundance and beauty in the garden. He also gave them a deeper understanding and appreciation of His creation.

{\color{red}11}.	Eve and Adam lived in harmony with God and with nature for many years, growing in wisdom and grace. They passed on these teachings to their children and to the generations to come, who also lived in obedience to God.

{\color{red}12}.	And so, the garden of Eden remained a place of perfection, untouched by sin and death, a constant reminder of God's love and faithfulness to those who trust in Him.
} \\
\bottomrule
\end{tabular}
};
\draw [rounded corners=.5em] (table.north west) rectangle (table.south east);
\end{small}
\end{tikzpicture}
\end{center}
\end{table*}

Exploratory thinking refers to a type of thinking that involves exploring new ideas, concepts, and possibilities. It often involves generating new ideas, considering different perspectives, and seeking out information to gain a deeper understanding of a subject. This type of thinking is often associated with creativity and innovation.

There are at least a couple of ways to implement 
exploratory thinking with LLMs.  
The first method is to use the {\em maieutics} strategy, which is one of the Socratic method \cite{SocraticIEEECCWC2023}.
The second method is to employ 
{\em counterfactual reasoning}. 

\subsubsection{Maieutics Method}
\hfill \break
The term ``maieutic'' comes from the Greek word ``maieutikos,'' which means midwife. The method of maieutics is based on the idea that the teacher's role is to help the student give birth to their own understanding of a subject, rather than to impart knowledge directly.

We use the example in Table~\ref{tab:Maieutics} to illustrate how GPT-3 starts from knowing nothing about the semantics of an idiom to being able to create new idioms expressing the same meaning with different metaphors. Initially, GPT-3 did not have a clear understanding of the Taiwanese idiom: ``planting gourd yields cucumber.'' The idiom conveys the meaning that when a farmer plants gourds, which are more valuable and have a higher market price, but unexpectedly harvests cucumbers, which have a lower market price, the farmer not only experiences disappointment but also suffers from financial loss.
GPT-3 initially could not understand the full meaning of the idiom.  After a couple of iterations, GPT-3 ``gives birth'' impressive new variations. 

\begin{table}[htbp]
\vspace{-.12in}
\caption{Example of Method Maieutics}
\vspace{-.12in}
\label{tab:Maieutics}
\begin{tikzpicture}
\begin{small}
\node (table) [inner sep=0pt] {
\begin{tabular}{p{1.2cm}|p{6.5cm}}
\toprule
Role & Dialogue \\
\midrule
Prompt & {Can you interpret the meaning of this Taiwanese idiom: "planting gourd yields cucumber?"}\\
GPT3 & {The idiom is often used to convey the idea that sometimes, despite your best efforts, things don't turn out the way you want them to..} \\
Prompt & {You are half right. It also implies that the price of gourd is much higher than the market price of cucumber. The farmer not only yields unexpected results, but also loses income! } \\

GPT3 & {It's an idiomatic expression that can be used to describe a situation in which someone works hard and invests a lot of resources into something, but the outcome is not only different from what was expected, but also less profitable. It conveys the idea of frustration and financial loss.} \\

Prompt & {Can you replace the two vegetables, gourd and cucumber, with two other plants and still maintain the meaning of the idiom?} \\

GPT3 & {These are three examples:} \\

& 1. "Planting strawberry yields raspberry," which
conveys the idea of disappointment and financial loss when someone plants more valuable and expensive strawberry but ends up getting less valuable and cheaper raspberry. \\
& 2. "Planting melon yields watermelon."... \\
& 3. "Planting apple yields pear."... \\
Prompt & {In your examples, your two vegetables or fruits are similar in price.  Please consider price when
formulating your idioms.} \\

GPT3 & {I apologize. These are three new examples:} \\
& 1. "Planting truffle yields mushroom."... \\
& 2. "Planting caviar yields roe."... \\
& 3. "Planting lobster yields crab."... \\

Prompt & {Great! In these examples you did consider values of the objects. But do you know that lobster and crab are not plants. They are both phylum arthropods and they live in water. They cannot be planted! }  \\

GPT3 & {... Sorry... Lobster and crab are not plants, and they cannot be planted. They are both members of the phylum arthropoda, and live in water. Instead of ``planting,'' other verbs should be appropriate for these animals, such as:} \\

& 1. "Planting truffle yields mushroom."... \\
& 2. "Harvesting caviar yields roe."... \\
& 3. "Catching lobster yields crab."... \\
\end{tabular}
};
\draw [rounded corners=.5em] (table.north west) rectangle (table.south east);
\end{small}
\end{tikzpicture}
\vspace{-.3in}
\end{table}

\subsubsection{Counterfactual Thinking}

Counterfactual thinking refers to imaginative thoughts about what might have been (``if only'' or ``what if'').
We have experimented with using the counterfactual technique to rewrite a chapter to connect the two greatest classical novels
\cite{CRITExtended2023}.
We have also asked GPT-3 to rewrite Genesis chapter 3 by 
prompting GPT-3 that: ``What if Adam and Eve refused the serpent to eat the fruit?''
Table~\ref{tab:genesis} presents GPT-3's creativity in completing Genesis chapter 6 after being given the prompt of ``what if.'' Notice that before prompting GPT-3 to start writing,
we convey our intent and ensure that GPT-3 has knowledge about the Bible as the context.

\section{Concluding Remarks}
\label{sec-conclusion}

The goal of this paper is to propose a framework for enhancing system-2 AI by developing a computational consciousness model capable of performing tasks that involve reasoning and decision-making while taking ethical considerations into account. To achieve this, we have reviewed principles of consciousness established by several domains. While the full understanding of consciousness is still incomplete, we are encouraged by the idea of functionalism, which suggests that the implementation of consciousness in machines need not strictly follow human anatomy.

The $\CCM$ model is a novel solution to the problem of current AI systems' limitations in terms of ethical considerations and emotional intelligence. By incorporating desired moral principles such as knowledge, fairness, beneficence, non-maleficence, empathy, adaptability, transparency, and critical and exploratory thinking abilities, $\CCM$ has the potential to create AI agents that combine both knowledge and compassion. 


We are actively exploring ways to link $\CCM$'s task priority setting and scheduling policy with an external reward system that is based on ethical considerations, in order to facilitate the management of tasks in an ethical manner. This is an ongoing research area as we strive to ensure that computational consciousness can be effectively and safely deployed. Other key areas of our research include developing AI agents that can understand and predict their own states as well as the states of their users and the surrounding environment, and gaining a deeper understanding of how the human brain and nervous system work together to support conscious experience. Techniques such as optogenetics \cite{deisseroth2005next,deisseroth2021projections} may provide new insights that can be applied to the development of computational consciousness.

\noindent
{\bf Update after the launch of GPT-4}: The performance of GPT-4 is impressive in performing traditional NLP tasks. Furthermore, the research conducted by \cite{bubeck2023sparks} indicates that ``GPT-4 can solve novel and difficult tasks that span mathematics, coding, vision, medicine, law, psychology and more, without needing any special prompting.'' GPT-4 also demonstrates common sense and a theory of mind, and ``it could reasonably be viewed as an early (yet still incomplete) version of an artificial general intelligence (AGI) system.''

During his talk at Stanford on April $5^{th}$, 2023 \cite{BubeckTalk4-5-2023} Sébastien Bubeck conveyed that planning is still a weakness of GPT-4, and although issues such as hallucination and safety have been improved, they still remain. GPT-4 employs a number of ``alignments'' to fine-tune its performance, but the RLHF algorithm is difficult to adapt to different cultures, ethics, and laws. The effects and side-effects of hundreds of alignments are unknown. In fact, GPT-4 acts as a black box, and it is difficult to determine whether it is telling the truth or what a user wants to hear. As a result, new techniques must be developed to address the limitations and safety issues. Our ongoing work involves applying $\CCM$ to mitigate some safety and ethical issues.

\section*{Acknowledgement}
I would like to thank my colleague Professor Monica Lam, as well as intern students Ethan Chang and Mason Wang, for their leadership and contributions to the design and development of the Noora prototype \cite{NooraStanford2022} since the summer of 2022 at Stanford University.

\renewcommand*{\bibfont}{\small}
\bibliographystyle{plainnat}
\bibliography{References}

\end{document}